\documentclass[preprint,showpacs,preprintnumbers,amsmath,amssymb]{revtex4}
\usepackage{bm}
\usepackage{graphicx}
\usepackage{floatflt,wrapfig}
\usepackage{fancyhdr}
\newcommand{\fcaption}[1]{\refstepcounter{figure} \footnotesize \baselineskip=12pt Fig.~\thefigure.~#1}

\begin{document}
\title{Theory of electric-field-induced spin accumulation and spin
current in the two-dimensional Rashba model}

\author{V.V. Bryksin}
\affiliation{Physical Technical Institute, Politekhnicheskaya 26,
194021 St. Petersburg, Russia}

\author{P. Kleinert}
\affiliation{Paul-Drude-Intitut f\"ur Festk\"orperelektronik,
Hausvogteiplatz 5-7, 10117 Berlin, Germany}

\date{\today}

\begin{abstract}
Based on the spin-density-matrix approach, both the
electric-field-induced spin accumulation and the spin current are
systematically studied for the two-dimensional Rashba model.
Eigenmodes of spin excitations give rise to resonances in the
frequency domain. Utilizing a general and physically well-founded
definition of the spin current, we obtain results that differ
remarkably from previous findings. It is shown that there is a
close relationship between the spin accumulation and the spin
current, which is due to the prescription of a quasi-chemical
potential and which does not result from a conservation law.
Physical ambiguities are removed that plagued former approaches
with respect to a spin-Hall current that is independent of the
electric field. For the clean Rashba model, the intrinsic
spin-Hall conductivity exhibits a logarithmic divergency in the
low-frequency regime.
\end{abstract}


\maketitle

\section{Introduction}
It has been anticipated that the spin degree of freedom of charge
carriers potentially provides additional functionality to
electronic devices. Recent progress aims at spintronic
applications that rely on the capability to manipulate electron
spin polarizations in nonmagnetic semiconductors. Of particular
interest are efficient injection mechanisms of spins in
semiconductors at room temperature. In this respect, the proposal
by Murakami et al. \cite{Murakami} and Sinova et al.
\cite{Sinova1} of generating dissipationless transverse spin
currents by a driving electric field has attracted considerable
attention. The observation of this spin-Hall effect has been
reported in recent experiments on GaAs and related materials
\cite{Kato_spin_hall,Wunderlich}. Many theoretical studies of this
effect
\cite{Dyakonov,Levitov,Rashba1,Inoue,Inoue1,Culcer,Schliemann69,%
Mishchenko,Erlingsson,Dimitrova,Zhang94,Khaetskii} focused on the
Rashba spin-orbit interaction since this type of spin-charge
coupling can be easily controlled by an electrical gate. The
original conclusion concerning the existence of an universal
intrinsic spin-Hall conductivity \cite{Sinova1} in clean systems
was reexamined in more detail by treating effects of the elastic
impurity scattering. Based on numerical \cite{Sheng,Nomura} and
analytical results derived from the Keldysh
\cite{Mishchenko,LiuLei} and Kubo
\cite{Dimitrova,Chalaev,Raimondi} formalism, it has been concluded
that vertex corrections lead to a vanishing zero-frequency
spin-Hall current in the thermodynamic limit of the linear Rashba
model. After long debates many researchers finally arrived at the
same conclusion so that there seems to be agreement now that the
intrinsic zero-frequency spin current is finite and universal for
a free two-dimensional electron gas but vanishes in impure systems
for an arbitrary ratio of spin splitting and the impurity
scattering rate.

Despite this consensus there are still challenging problems to be
addressed referring to a proper definition of the spin current.
This issue has recently been treated by Zhang et al. \cite{PZhang}
and Sugimoto et al. \cite{Sugimoto}. The authors pointed out that
the "conventional" spin current, which is defined as the product
of spin and velocity operators, loses its physical foundation when
the spin-orbit coupling is present. The main deficiency of this
definition is related to the absence of a conservation law of
spins. Therefore, a physically motivated definition was suggested
\cite{PZhang,Sugimoto} that relates the spin current to the time
derivative of the spin displacement. It was argued that under
quite general conditions this effective spin current satisfies the
continuity equation so that it is measurable as a spin
accumulation. While for the charge transport both definitions
completely agree to each other, there is a remarkable discrepancy
between them with respect to the spin current due to the torque
dipole contribution \cite{PZhang}. Results for the spin Hall
conductivity derived from this physically motivated definition of
the spin current remarkably differ from previous findings based on
the "conventional" definition. Unfortunately, the authors did not
apply direct perturbational techniques to calculate this spin-Hall
current. Rather, they worked out a special calculational schema
for the determination of the torque dipole density, which they
used to complement the "conventional" spin-Hall current to a
conserved quantity. In view of the long and successful history to
derive the current from the time derivative of the dipole moment
\cite{BoBuch}, we prefer the application of the standard
procedure. What is needed in this approach is nothing but the
density matrix, which is determined from quantum-kinetic
equations. Starting from the physical definition of the spin-Hall
current, we calculate the frequency dependent spin polarization
and spin-Hall conductivity by analytically solving the kinetic
equations for the spin-density matrix. The obtained results do not
agree with previous findings deduced from the "conventional"
definition of the spin-Hall current. Furthermore, physical
inconsistencies with respect to a spin-Hall current component that
is independent of the electric field as well as the relationship
between the spin accumulation and the spin current are puzzled
out. For the spin-Hall conductivity of a clean two-dimensional
electron gas with Rashba type spin-orbit coupling, our exact
calculation does not reproduce any universal value but yields a
logarithmic divergency in the low frequency limit. In addition,
the frequency dependent spin-Hall conductivity, which is obtained
by a controlled perturbational approach, differs even
qualitatively from previous results \cite{Mishchenko}.

\section{The kinetic equations}
We consider a two-dimensional electron gas in the presence of
Rashba spin-orbit interaction with amplitude $\alpha$. The system
with an applied in-plane electric field $\vec{\bm{E}}$ (which is
oriented along the $x$ axis) is described by the Hamiltonian
\begin{eqnarray}
H_{0}&=&\sum_{\bm{k},\lambda }a_{\bm{k}\lambda }^{\dag}\left[ \varepsilon_{%
\bm{k}}-\varepsilon _{F}\right] a_{%
\bm{k}\lambda }-\sum_{\bm{k},\lambda ,\lambda ^{\prime }}\left(
\hbar\vec{\bm{\omega}}_{
\bm{k}} \cdot \vec{\bm{\sigma }}_{\lambda \lambda ^{\prime }}\right) a_{\bm{k}%
\lambda }^{\dag}a_{\bm{k}\lambda ^{\prime }}\nonumber\\
&-&e\vec{\bm{E}} \sum_{\bm{k},\lambda}\left. \nabla_{\bm \kappa}
a^{\dag}_{\bm{k}-\frac{\bm{\kappa}}{2}\lambda}a_{\bm{k}+
\frac{\bm{\kappa}}{2}\lambda}\right|_{\bm{\kappa}=\bm{0}},
\label{Hamil}
\end{eqnarray}
where we introduced the abbreviations
\begin{equation}
\varepsilon_{\bm{k}}=\frac{\hbar^2\bm{k}^2}{2m},\quad
\vec{\bm{\omega}}_{
\bm{k}}=\frac{\hbar}{m}(\bm{K}\times\bm{k}),\quad
\bm{K}=\frac{m\alpha}{\hbar^2}\vec{\bm{e}}_z . \label{eq2}
\end{equation}
$m$, $\varepsilon_F$, and $\vec{\bm{\sigma}}$ denote the effective
mass, the Fermi energy, and the vector of Pauli matrices,
respectively. $a_{\bm{k}\lambda}^{\dag}$ and $a_{\bm{k}\lambda}$
are creation and annihilation operators with quasimomentum
$\bm{k}=(k_x,k_y,0)$ and spin $\lambda$. We are going to calculate
the time-dependent density matrix
\begin{equation}
f_{\lambda^{\prime}}^{\lambda}(\bm{k},\bm{k}^{\prime}\mid
t)=\langle
a^{\dag}_{\bm{k}\lambda}a_{\bm{k}^{\prime}\lambda^{\prime}}\rangle_t,
\label{eq3}
\end{equation}
which is more conveniently expressed by its physical
representation
\begin{equation}
f(\bm{k},\bm{\kappa}\mid
t)=\sum\limits_{\lambda}f_{\lambda}^{\lambda}(\bm{k},\bm{\kappa}\mid
t),\quad \vec{\bm{f}}(\bm{k},\bm{\kappa}\mid
t)=\sum\limits_{\lambda,\lambda^{\prime}}f_{\lambda^{\prime}}^{\lambda}(\bm{k},\bm{\kappa}\mid
t)\vec{\bm{\sigma}}_{\lambda,\lambda^{\prime}},%
\label{eq4}
\end{equation}
in the $\bm{k}$, $\bm{\kappa}$ space, where $\bm{k}\rightarrow
\bm{k}+\bm{\kappa}/2$ and $\bm{k}^{\prime}\rightarrow
\bm{k}-\bm{\kappa}/2$. $\bm{\kappa}$ refers to a possible
inhomogeneity of the charge and/or spin distribution. Using the
Liouville equation, the quantum-kinetic equations for the
components of the density matrix are straightforwardly derived.
From the result
\begin{eqnarray}
&&\frac{\partial f}{\partial
t}-\frac{i\hbar}{m}(\bm{\kappa}\cdot\bm{k})f-\frac{i\hbar}{m}\bm{K}(\vec{\bm{f}}\times\bm{\kappa})
+\frac{e\vec{\bm{ E}}}{\hbar}\nabla_{\bm{k}}f
\\
&&=\sum\limits_{\lambda,\lambda_1,\lambda_2}\sum\limits_{\bm{k}^{\prime}}\biggl\{
f_{\lambda_2}^{\lambda_1}(\bm{k}^{\prime},\bm{\kappa}\mid
t)W_{\lambda_2 \lambda}^{\lambda_1
\lambda}(\bm{k}^{\prime},\bm{k},\bm{\kappa})
-f_{\lambda_2}^{\lambda_1}(\bm{k},\bm{\kappa}\mid t)W_{\lambda_2
\lambda}^{\lambda_1 \lambda}(\bm{k},\bm{k}^{\prime},\bm{\kappa})
\biggl\}\equiv I, \nonumber%
\label{eq5}
\end{eqnarray}
\begin{eqnarray}
&&\frac{\partial\vec{\bm{f}}}{\partial
t}-\frac{i\hbar}{m}(\bm{\kappa}\cdot\bm{k})\vec{\bm{f}}+2(\vec{\bm{\omega}}\times\vec{\bm{f}})
+\frac{i\hbar}{m}(\bm{K}\times\bm{\kappa})f+\frac{e\vec{\bm{
E}}}{\hbar}\nabla_{\bm{k}}\vec{\bm{f}}\\
&&=\sum\limits_{\lambda_1,\lambda_2}\sum\limits_{\lambda_3,\lambda_4}
\sum\limits_{\bm{k}^{\prime}}\biggl\{f_{\lambda_2}^{\lambda_1}(\bm{k}^{\prime},\bm{\kappa}\mid
t)W_{\lambda_2 \lambda_4}^{\lambda_1
\lambda_3}(\bm{k}^{\prime},\bm{k},\bm{\kappa})
-f_{\lambda_2}^{\lambda_1}(\bm{k},\bm{\kappa}\mid t)W_{\lambda_2
\lambda_4}^{\lambda_1
\lambda_3}(\bm{k},\bm{k}^{\prime},\bm{\kappa})\biggl\}\vec{\bm{\sigma}}_{\lambda_3\lambda_4}
\equiv \vec{\bm{I}} \nonumber,%
\label{eq6}
\end{eqnarray}
it is concluded that not only scattering but also any
inhomogeneity ($\bm{\kappa}\ne \bm{0}$) couples the charge ($f$)
and spin ($\vec{\bm{f}}$) degrees of freedom to each other.
Consequently, any accumulation of charges induces a spin response
and vice versa. The left-hand side of these equations was derived
and discussed in Ref. \cite{Mishchenko03}. The scattering
probabilities $W_{\lambda_2 \lambda_4}^{\lambda_1 \lambda_2}$ on
the right hand sides of Eqs.~(5) and (6) comprise both elastic and
inelastic scattering-in and scattering-out contributions, which
satisfy a sum rule. We shall restrict the consideration to elastic
scattering described by the Hamiltonian
\begin{equation}
H_{int}=u\sum\limits_{\bm{k},
\bm{k}^{\prime}}\sum\limits_{\lambda}a_{\bm{k}\lambda}^{\dag}a_{\bm{k}^{\prime}\lambda},%
\label{eq7}
\end{equation}
with $u$ denoting the magnitude of the short-range impurity
potential. For time-dependent phenomena, we prefer the treatment
of the Laplace transformed kinetic Eqs.~(5) and (6). Within the
Born approximation, we obtain for the scattering probabilities the
exact result
\begin{eqnarray}
&&W_{\lambda _2\lambda _4}^{\lambda _1\lambda _3}\left( \bm{k}^{\prime },\bm{k%
},\bm{\kappa}\mid s\right) =\frac{u^2}{\hbar ^2}\int_0^\infty
dt\exp \left[ -st+\frac i\hbar \left( \varepsilon_{\bm{k}^{\prime
}-\bm{\kappa}/2}-\varepsilon_{\bm{k}+\bm{\kappa}/2}\right)
t\right]\nonumber\\
&&\times \left[ \cos \left(
\omega _{\bm{k}^{\prime }-\bm{\kappa}/2}t\right) \delta _{\lambda _1\lambda _3}-i\frac{\vec{{\bm %
\sigma }}_{\lambda _1\lambda
_3}\cdot{\vec{\bm{\omega}}}_{\bm{k}^{\prime
}-\bm{\kappa}/2}}{\omega _{\bm{k}^{\prime }-\bm{\kappa}/2}}\sin
\left( \omega _{\bm{k}^{\prime}-\bm{\kappa}/2}t\right)
\right]\nonumber\\
&&\times \left[ \cos \left( \omega _{\bm{k}+\bm{\kappa}/2}t\right) \delta _{\lambda _4\lambda _2}+i%
\frac{{\vec{\bm{\sigma}}}_{\lambda _4\lambda
_2}\cdot{\vec{\bm{\omega}}}_{\bm{k}+\bm{\kappa}/2}}{\omega
_{\bm{k}+\bm{\kappa}/2}}\sin \left( \omega
_{\bm{k}+\bm{\kappa}/2}t\right) \right] +\bm{k}\rightleftarrows
\bm{k}^{\prime },%
\label{eq8}
\end{eqnarray}
with $s$ denoting the variable of the Laplace transformation. As
usual, it is assumed that corrections due to the $\bm{\kappa}$
expansion of the scattering probabilities are small compared to
corresponding contributions on the left-hand side of the kinetic
equations. Furthermore, for weak spin-orbit coupling, we may
restrict to the lowest-order contributions in $\omega_{\bm{k}}t$.
Adopting these approximations, the collision integrals are
expressed by
\begin{eqnarray}
I&=&\frac{1}{\tau}(\overline{f}-f)%
-\frac{\hbar\vec{\bm{\omega}}_{\bm{k}}}{\tau}\frac{\partial^2}{\partial
\varepsilon_{\bm{k}}^2}\overline{\hbar\vec{\bm{\omega}}_{\bm{k}}f}%
+\frac{1}{\tau}\frac{\partial}{\partial\varepsilon_{\bm{k}}}
\overline{\hbar\vec{\bm{\omega}}_{\bm{k}}\cdot\vec{\bm{f}}}\label{e10}\\%
&-&\frac{\hbar\vec{\bm{\omega}}_{\bm{k}}}{\tau}\frac{\partial}{\partial\varepsilon_{\bm{k}}}%
\overline{\vec{\bm{f}}}+4\hbar u^2\sum\limits_{\bm{k}^{\prime}}%
\vec{\bm{f}}(\bm{k}^{\prime})\frac{\vec{\bm{\omega}}_{\bm{k}^{\prime}}\times%
\vec{\bm{\omega}}_{\bm{k}}}{(\varepsilon_{\bm{k}}-\varepsilon_{\bm{k}^{\prime}})^3},%
\nonumber
\end{eqnarray}
\begin{eqnarray}
\vec{\bm{I}}&=&\frac{1}{\tau}(\overline{\vec{\bm{f}}}-\vec{\bm{f}})%
+\frac{\hbar\omega_{\bm{k}}}{\tau}\frac{\partial^2}{\partial
\varepsilon_{\bm{k}}^2}\overline{\hbar\omega_{\bm{k}}\vec{\bm{f}}}%
+\frac{1}{\tau}\frac{\partial}{\partial\varepsilon_{\bm{k}}}
\overline{\hbar\vec{\bm{\omega}}_{\bm{k}} f}
-\frac{\hbar\vec{\bm{\omega}}_{\bm{k}}}{\tau}\frac{\partial}{\partial\varepsilon_{\bm{k}}}%
\overline{f}\label{e11}\\%
&+&%
\frac{\hbar}{\tau}\left[\vec{\bm{\omega}}_{\bm{k}}\times\left(\frac{\partial^2}%
{\partial\varepsilon_{\bm{k}}^2}%
\hbar\vec{\bm{\omega}}_{\bm{k}}\times\vec{\bm{f}}\right)\right]%
-2u^2\sum\limits_{\bm{k}^{\prime}}\frac{\left[(\vec{\bm{\omega}}_{\bm{k}}+%
\vec{\bm{\omega}}_{\bm{k}^{\prime}})\times\vec{\bm{f}}(\bm{k}^{\prime})\right]}%
{(\varepsilon_{\bm{k}}-\varepsilon_{\bm{k}^{\prime}})^2}%
+4\hbar u^2\sum\limits_{\bm{k}^{\prime}}f(\bm{k}^{\prime})%
\frac{\vec{\bm{\omega}}_{\bm{k}^{\prime}}\times\vec{\bm{\omega}}_{\bm{k}}}%
{(\varepsilon_{\bm{k}^{\prime}}-\varepsilon_{\bm{k}})^3},
\nonumber
\end{eqnarray}
where the scattering time $\tau$ is calculated from
\begin{equation}
\frac{1}{\tau}=\frac{2\pi u^2}{\hbar}\sum\limits_{\bm{k}^{\prime}}
\delta(\varepsilon_{\bm{k}^{\prime}}-\varepsilon_{\bm{k}}).%
\label{deftau}
\end{equation}
$\overline{f}(\bm{k})$ means an average over the angle of the
vector $\bm{k}$. In the Eqs.~(\ref{e10}) and (\ref{e11})
corrections appear, which result from virtual transitions that are
not linked to the scattering time $\tau$. The treatment of these
contributions as well as higher-order corrections in the
spin-orbit coupling $\alpha$ goes beyond the scope of this paper.
Restricting to the lowest-order scattering contributions, assuming
an initial thermodynamic equilibrium state, and focusing on weak
spin-orbit coupling so that $\hbar^2K/(m\varepsilon_{\bm{k}})\ll
1$, the kinetic equations are expressed by
\begin{equation}
sf-\frac{i\hbar}{m}(\bm{\kappa}\cdot\bm{k})f-\frac{i\hbar}{m}\bm{K}(\vec{\bm{f}}\times\bm{\kappa})
+\frac{e\vec{\bm{ E}}}{\hbar}\nabla_{\bm{k}}f
=\frac{1}{\tau}(\overline{f}-f)+n(\varepsilon_{\bm{k}}),
\label{kin1}
\end{equation}
\begin{eqnarray}
&&s\vec{\bm{f}}+2(\vec{\bm{\omega}}_{\bm{k}}\times\vec{\bm{f}})
-\frac{i\hbar}{m}(\bm{\kappa}\cdot\bm{k})\vec{\bm{f}}
+\frac{i\hbar}{m}(\bm{K}\times\bm{\kappa})f+\frac{e\vec{\bm{
E}}}{\hbar}\nabla_{\bm{k}}\vec{\bm{f}}\nonumber\\
&&=\frac{1}{\tau}(\overline{\vec{\bm{f}}}-\vec{\bm{f}})+\frac{1}{\tau}
\frac{\partial}{\partial\varepsilon_{\bm{k}}}
\overline{f\hbar\vec{\bm{\omega}}_{\bm{k}}}-\frac{\hbar\vec{\bm{\omega}}_{\bm{k}}}{\tau}
\frac{\partial}{\partial\varepsilon_{\bm{k}}} \overline{f}
-\hbar\vec{\bm{\omega}}_{\bm{k}}\frac{\partial
n(\varepsilon_{\bm{k}})}{\partial \varepsilon_{\bm{k}}},
\label{kin2}
\end{eqnarray}
where $n(\varepsilon_{\bm{k}})$ denotes the initial equilibrium
charge density.

\section{General expressions for the charge and spin currents}
The Laplace-transformed density matrix
$\widehat{f}(\bm{k},\bm{\kappa}\mid s)=\{f(\bm{k},\bm{\kappa}\mid
s),\vec{\bm{f}}(\bm{k},\bm{\kappa}\mid s)\}$ contains all the
information needed to determine all kinetic observables. In
particular, the quantity
\begin{equation}
f(s)=\sum\limits_{\bm{k}}f(\bm{k},\bm{\kappa}\mid
s)\mid_{\bm{\kappa}=\bm{0}}%
\label{dd1}
\end{equation}
represents nothing but the conserved charge density. Exactly in
the same way, the total magnetic moment is calculated from
\begin{equation}
\vec{\bm{f}}(s)=
\sum\limits_{\bm{k}}\vec{\bm{f}}(\bm{k},\bm{\kappa}\mid
s)\mid_{\bm{\kappa}=\bm{0}}.%
\label{dd2}
\end{equation}
When an electric field is applied to the system, these quantities
become time dependent and allow the treatment of relaxation
processes.

Other quantities of interest are the current of charge carriers
and spins, which are obtained from
\begin{equation}
\vec{\bm{j}}(s)=-ise\sum\limits_{\bm{k}}\nabla_{\bm{\kappa}}%
\vec{\bm{f}}(\bm{k},\bm{\kappa}\mid s)\mid_{\bm{\kappa}=\bm{0}},%
\label{charge}
\end{equation}
and
\begin{equation}
\widehat{\bm{j}}^s(s)=-is\frac{1}{2}\sum\limits_{\bm{k}}\nabla_{\bm{\kappa}}\otimes%
\vec{\bm{f}}(\bm{k},\bm{\kappa}\mid s)\mid_{\bm{\kappa}=\bm{0}},
\label{spinc}
\end{equation}
respectively.  In Eq.~(\ref{spinc}), $\otimes$ denotes the dyadic
product. These definitions are fundamental and sufficiently
general. In the time domain, the spatial versions of these
equations describe the temporal evolution of the carrier or spin
displacement, i.e., the center-of-mass velocity of the wave
packet. The expression for the charge-carrier current in
Eq.~(\ref{charge}) is also applicable for systems without any
spatial dispersion. If the interaction Hamiltonian that describes
elastic or inelastic scattering commutes with the dipole operator
of the carriers than the definition in Eq.~(\ref{charge}) becomes
completely equivalent to
\begin{equation}
j_i(t)=\frac{e}{\hbar}\sum\limits_{\bm{k}}\frac{\partial\varepsilon_{\bm{k}}}
{\partial k_i}f(\bm{k}\mid
t)+\frac{e}{\hbar}\sum\limits_{\bm{k},j}
\frac{\partial^2\varepsilon_{\bm{k}}}{\partial k_i\partial
k_j}\left(\bm{K}\times\vec{\bm{f}}(\bm{k}\mid t)\right)_j,%
\label{spincH}
\end{equation}
in which the $\bm{\kappa}$ dependence does no longer occur. We
want to stress that the situation for the spin transport is
completely different and more subtle. For a system without any
spatial dispersion, it is in general not possible to express the
spin current as defined in Eq.~(\ref{spinc}) by the density matrix
$\widehat{f}(\bm{k},\bm{\kappa}\mid s)\mid_{\bm{\kappa}=\bm{0}}$
alone. What is really needed is the $\bm{\kappa}$ derivative of
this function at $\bm{\kappa}=\bm{0}$. Notwithstanding this fact,
most researchers defined the spin current $\widehat{j}^s(s)$ as
the symmetrized product of the spin and velocity operators
$\{\widehat{\vec{\sigma}}_i,{\vec{v}}_k\}_+/4$, where
$\widehat{\vec{v}}=\nabla_{\bm{k}}H_0/\hbar$ and $H_0$ is obtained
from Eq.~(\ref{Hamil}) for $\vec{\bm{E}}=\bm{0}$. For the Rashba
model, this definition of the spin current takes the for
\begin{equation}
\widehat{j}^s(s)=\frac{1}{2\hbar}\sum\limits_{\bm{k}}\nabla_{\bm{k}}
\varepsilon_{\bm{k}}\otimes \vec{\bm{f}}(\bm{k},\bm{\kappa}\mid
s)\mid_{\bm{\kappa}=\bm{0}}. \label{conv}
\end{equation}
As shown below, the definition in Eq.~(\ref{conv}) leads to a
stationary spin-Hall current that does not depend on the electric
field. In addition, recent studies \cite{PZhang,Sugimoto} clearly
demonstrated that results derived from the Eqs.~(\ref{spinc}) and
(\ref{conv}) considerably differ from each other and that the
measurable quantity is related to Eq.~(\ref{spinc}). Furthermore,
we note that the diffusion tensor is likewise obtained from
derivatives of the density matrix
$\widehat{f}(\bm{k},\bm{\kappa}\mid s)$ with respect to $\kappa_i$
at $\bm{\kappa}=\bm{0}$. Although the extension of our approach to
the treatment of the diffusion coefficient is straightforward, we
want to confine ourselves to the analysis of spin currents.

The definitions of measurable quantities in Eqs.~(14) to (18)
already set up our calculational schema as an iteration with
respect to $\bm{\kappa}$. As we restrict ourselves to elastic
scattering, the spectral functions can be calculated for a given
energy over which one finally integrates. This procedure is
applicable only in the linear response regime, where carrier
heating and energy relaxation due to nonlinear field effects do
not play an essential role. Therefore, we treat only first-order
corrections in the electric field. The analytic solution of the
kinetic Eqs.~(\ref{kin1}) and (\ref{kin2}), which is
straightforward but cumbersome is presented in the Appendix.

\section{Spin accumulation}
To begin with the solution of the kinetic equation, which is
derived in the Appendix, is used for the calculation of the spin
accumulation. From Eq.~(\ref{mvf0E}), we obtain for the vector of
the field-induced components of the spin-density matrix
\begin{equation}
\vec{\bm{f}}(s)=-\frac{e\hbar}{ms}(\bm{K}\times\vec{\bm{E}})\sum\limits_{\bm{k}}
n^{\prime}(\varepsilon_{\bm{k}})\frac{2\tau\omega_{\bm{k}}^2}{\sigma^2
s\tau+2\omega_{\bm{k}}^2(2s\tau+1)},%
\label{Edel}
\end{equation}
with $\sigma=s+1/\tau$. According to Eq.~(\ref{Edel}), the
in-plane spin accumulation is calculated by a $\bm{k}$ integral
over poles. Under the condition $\omega_{\bm{k}}\tau >1$, when one
expects sharp resonances, the positions of these poles are given
by
\begin{equation}
(s\tau)_1=-\frac{1}{2},\quad (s\tau)_{2,3}=-\frac{3}{4}\pm
2i\omega_{\bm{k}}\tau.%
\label{Wurzel}
\end{equation}
The resonance is most pronounced at zero temperature ($T=0$) and
when $\omega_{\bm{k}}\tau\gg 1$. Depending on the Rashba coupling
constant $\alpha$ and on the carrier density, the resonance at
$2\omega_{{k}_F}$ (${k}_F$ denotes the Fermi wave vector) is
located in the THz regime. Switching from the Laplace to the
frequency domain ($s\rightarrow -i\omega$) and considering zero
temperature, we obtain from Eq.~(\ref{Edel})
\begin{equation}
f^y(\omega)=eE\tau\frac{K}{\pi\hbar}
\frac{2i\omega_{k_F}^2}{\omega\tau\left[4\omega_{k_F}^2-(\omega+i/\tau)^2\right]
+2i\omega_{k_F}^2}, \label{spinaccu}
\end{equation}
with $\omega_{k_F}=\hbar Kk_F/m$. It is a striking coincidence
that the denominator in this expression for the spin accumulation
completely agrees with the denominator in the spin-Hall
conductivity calculated by Mishchenko et al. \cite{Mishchenko}.
The associated spin excitation gives rise to resonances in the
spin accumulation. An example is shown in Fig.~1 for some
spin-orbit coupling parameters $\omega_{k_F}\tau$. According to
the
\begin{figure}[!t]
\begin{minipage}[t]{7.5cm}
\centerline{\includegraphics*[width=7.5cm]{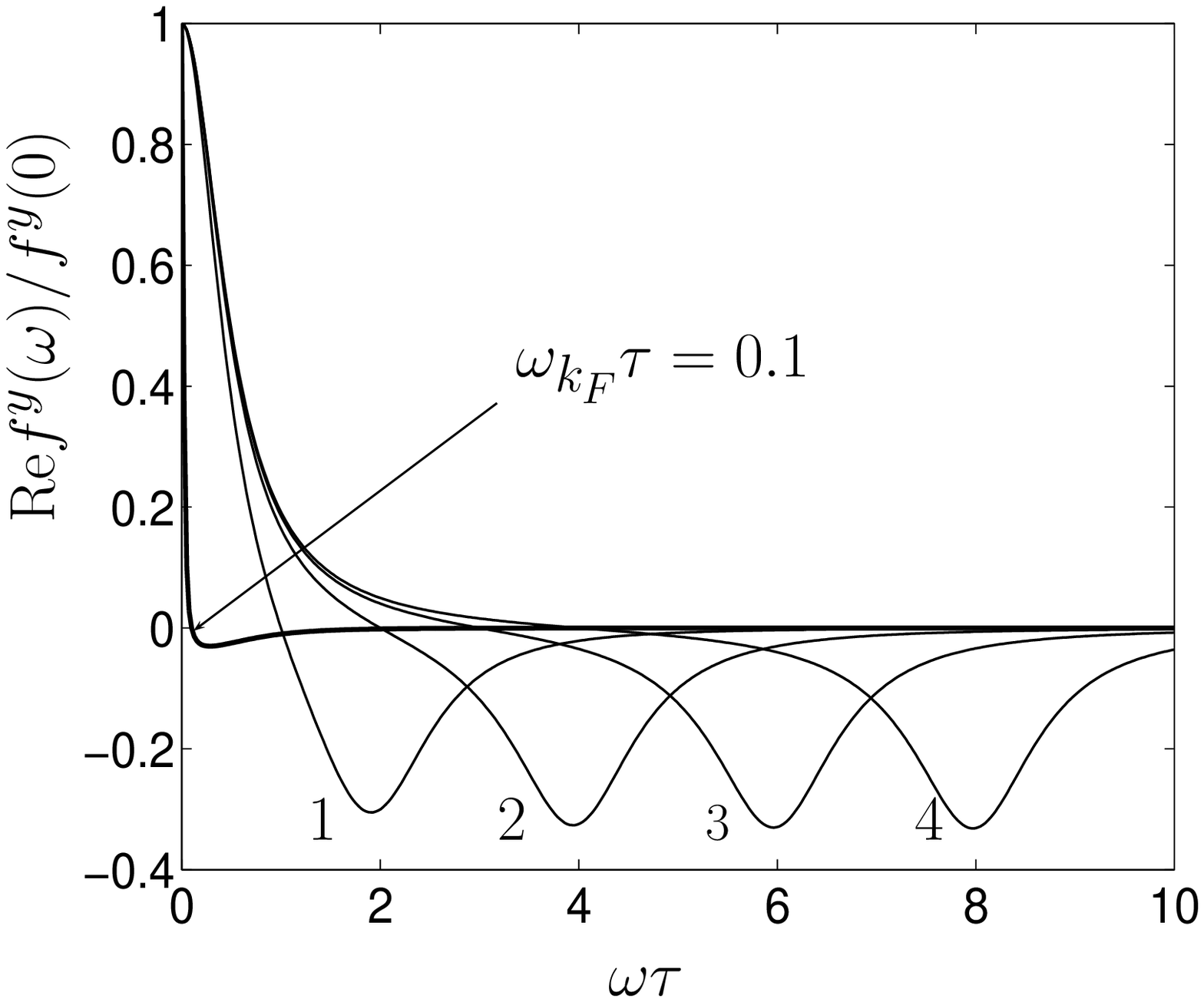}}
\fcaption{Frequency dependence of the normalized non-vanishing
component of the spin density matrix $f^y(\omega)$ for
$\omega_{k_F}\tau=0.1$, $1$, $2$, $3$, and $4$ with $f^y(0)=eE\tau
K/(\pi\hbar)$. \label{abb1}}
\end{minipage}
\hfill
\begin{minipage}[t]{7.5cm}
\centerline{\includegraphics*[width=7.5cm]{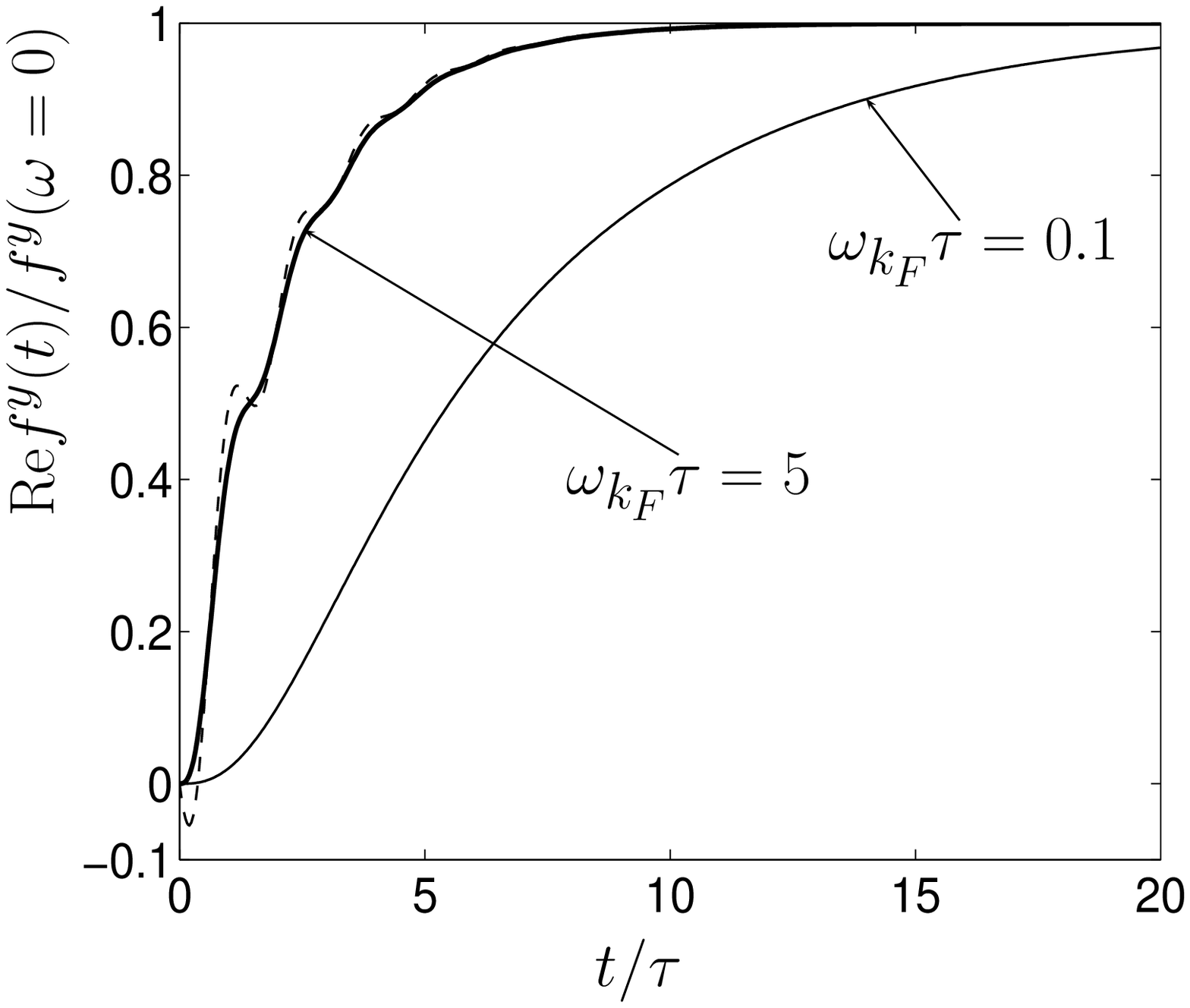}} \fcaption{Time
dependence of the normalized non-vanishing component of the spin
density matrix $f^y(t)$ for $\omega_{k_F}\tau=0.1$, $5$. The
dashed line is calculated from Eq.~(\ref{e26}). \label{abb2}}
\end{minipage}
\end{figure}
$\bm{k}$-integral in Eq.~(\ref{Edel}), the sharp peak is
increasingly washed out with increasing temperature. The
zero-frequency limit of the spin accumulation $f^y(\omega =0)$
agrees with the result published by Edelstein \cite{Edelstein}.

Applying an inverse Laplace transformation to Eq.~(\ref{Edel}),
the time evolution of the spin accumulation after the electric
field is switched on can be determined. Numerical results are
shown in Fig.~2 (solid lines) and compared with the following
analytical solution (dashed line)
\begin{equation}
f^y(t)=eE\tau\frac{K}{\pi\hbar}\left[1-\exp\left(-\frac{t}{2\tau}\right)
-\frac{\sin(2\omega_{k_F}t)}{2\omega_{k_F}\tau}\exp\left(-\frac{3t}{4\tau}\right)
\right],\label{e26}
\end{equation}
valid under the condition $\omega_{k_F}\tau \gg 1$. As expected,
for systems with a weak Rashba spin-orbit coupling, the steady
state spin accumulation is only reached after a sufficiently long
time. With decreasing $\omega_{k_F}\tau$, weak oscillations are
strongly suppressed.%

\section{Charge current}
According to the consideration in Section III, the longitudinal
current is calculated from the time derivative of the dipole
operator [cf. Eq.~(\ref{charge})]. Taking into account
Eq.~(\ref{mfkE}) in the Appendix, we immediately obtain for the
longitudinal charge current
\begin{equation}
j_x(s)=-4\frac{e^2E}{\sigma sm}\sum\limits_{\bm{k}}\varepsilon
n^{\prime}+2\frac{e^2E\tau}{s}\left(\frac{\hbar
K}{m}\right)^2\sum\limits_{\bm{k}}%
\frac{\omega_{\bm{k}}^2n^{\prime}}{\sigma^2s\tau+2\omega_{\bm{k}}^2(2s\tau+1)},%
\label{chargecc}
\end{equation}
which can be used to calculate the frequency response at zero
temperature ($n^{\prime}$ denotes the derivative
$dn(\varepsilon_{\bm{k}})/d\varepsilon_{\bm{k}}$). The result in
the frequency domain
\begin{equation}
j_x(\omega)=\frac{2\varepsilon_F\tau}{\pi
\hbar^2}\frac{e^2E}{1-i\omega\tau}-e^2E\tau\frac{K^2}{\pi
m}\frac{2i\omega_{k_F}^2}{\omega\tau\left[4\omega_{k_F}^2
-(\omega+i/\tau)^2\right]+2i\omega_{k_F}^2}, \label{ccurrent}
\end{equation}
is composed of two contributions. The first one is expressed by
the well-known Drude conductivity. The second one is due to the
spin-orbit interaction and exhibits the same resonant denominator
as the spin accumulation in Eq.~(\ref{spinaccu}). At $\omega=0$,
Eq.~(\ref{ccurrent}) reproduces the result published in Ref.
\cite{Inoue}. The measurement of the resonant longitudinal charge
current contribution allows the determination of the Rashba
coupling constant $\alpha$. The alternative formulation in
Eq.~(\ref{spincH}) exactly reproduces Eq.~(\ref{chargecc}).

The dynamical charge-Hall current in the Rashba split system has
been studied recently \cite{ZhangMa}. The appearance of this
effect is related to the $\bm{k}$ dependence of the scattering
matrix elements. As this dependence is not taken into account in
our approach, we obtain a vanishing charge Hall conductivity.

\section{Spin-Hall effect}
To introduce the spin-Hall effect, let us first shortly
recapitulate the main findings derived in the literature based on
the "conventional" definition of the spin current in
Eq.~(\ref{conv}). The treatment of the spin current within this
framework reveals an anomalousness, which in our opinion was not
duly noticed by many researchers. The approach predicts a
non-vanishing stationary spin-Hall current that is independent of
the electric field. Indeed, inserting Eq.~(\ref{f00}) into the
expression in Eq.~(\ref{conv}) for the spin current, we obtain
\begin{equation}
j_y^x(s)=\frac{\hbar
K}{2ms}\sum\limits_{\bm{k}}\varepsilon_{\bm{k}}n^{\prime}(\varepsilon_{\bm{k}}),
\label{zeroH}
\end{equation}
which leads to the constant $x$ component of the spin-Hall current
$j_y^x(\omega)=-K\varepsilon_F/(2\pi\hbar)$ in the frequency
domain at zero temperature. The absence of any resonance indicates
that there is no relationship between this fictitious spin-Hall
current and the spin accumulation in Eq.~(\ref{spinaccu}).

The field-induced spin-Hall current is calculated from its
definition in Eq.~(\ref{conv}) and by taking into account
Eq.~(\ref{mvf0E}). For zero temperature, we obtain for the
frequency-dependent spin-Hall current
\begin{equation}
j_y^z(\omega)=-\frac{eE}{2\pi\hbar}\,\frac{\omega\tau\omega_{k_F}^2}
{\omega\tau\left[4\omega_{k_F}^2-(\omega+i/\tau)^2\right]+2i\omega_{k_F}^2},
\label{Halperin}
\end{equation}
which was recently derived by applying the Keldysh approach
\cite{Mishchenko}. Here, the same resonant denominator appears as
in the longitudinal charge current [Eq.~(\ref{ccurrent})] and the
spin accumulation [eq.~(\ref{spinaccu})]. For a free electron gas
($\tau\rightarrow \infty$), Eq.~(\ref{Halperin}) simplifies to
\begin{equation}
j_y^z(\omega)=-\frac{eE}{2\pi\hbar}\frac{\omega_{k_F}^2}{4\omega_{k_F}^2-\omega^2},%
\label{Erli}
\end{equation}
which was previously obtained by Erlingsson et al.
\cite{Erlingsson}. Finally the steady state spin-Hall current
($\omega\rightarrow 0$) of the clean Rashba model is given by the
universal value \cite{Murakami,Sinova1}
\begin{equation}
j_y^z(\omega=0)=-\frac{eE}{8\pi\hbar}.%
\label{Sinov}
\end{equation}
Although many authors confirmed these results, there remain some
reservations. First of all, the definition of the "conventional"
spin-Hall current led to considerable confusion and to serious
doubts on its experimental relevance
\cite{Zhang94,PZhang,Sugimoto}. The main difficulty results from
the fact that the spin is not a conserved quantity. As it has been
claimed recently, a proper definition of the spin current requires
a careful analysis of the torque density. It is assumed that this
quantity may complement the above fictitious current to a
conserved spin current. The prerequisite for such a construction
is given, when the averaged spin-torque density vanishes in the
bulk \cite{PZhang}. It has been argued that this condition is
fortunately satisfied for many spin models treated in the
literature. This conserved spin current has the advantage that it
can be measured via the spin accumulation to which it is related
by the continuity equation. It has been pointed out that this
quantity is straightforwardly calculated from the time derivative
of the spin displacement (Eq.~(5) in Ref. \cite{PZhang}). A firm
foundation of this approach provides the definition of the spin
current in Eq.~(\ref{spinc}) and the kinetic equation treated in
Section II and solved in the Appendix.

Based on the physically motivated definition of the spin current
in Eq.~(\ref{spinc}) and using the analytical solutions of the
kinetic equations presented in the Appendix, we shall restart the
study of the spin-Hall current of the Rashba model. First, it is
noted that we also get a spin-Hall current contribution that is
independent of the electric field. From Eq.~(\ref{spinc}) and
(\ref{mvfk0}), we obtain a result
\begin{equation}
j_y^x(s)=-\frac{\hbar
K}{m}\sum\limits_{\bm{k}}n(\varepsilon_{\bm{k}})%
\frac{\omega_{\bm{k}}^2\tau}{\sigma^2
s\tau+2\omega_{\bm{k}}^2(2s\tau+1)}%
\label{zerowir}
\end{equation}
that differs from Eq.~(\ref{zeroH}) in many respects. First of
all, this current contribution is closely related to the spin
accumulation in Eq.~(\ref{Edel}). A first evidence for this
conclusion is the appearance of the same resonant denominator.
However, this relationship goes even deeper and has a firm
physical foundation, which becomes obvious by comparing the
analytical solution for $\vec{\bm{f}}_{0\bm{E}}$
[Eq.~(\ref{vf0E})] with its counterpart for
$\vec{\bm{f}}_{\bm{\kappa}0}$ [Eq.~(\ref{vfk0})]. One solution is
obtained from the other one by the replacement
$\bm{\kappa}\rightarrow -ieE\vec{\bm{e}}_x\partial_{\varepsilon}$,
where the derivative with respect to the energy applies to the
charge density $n$. Transforming back this replacement to the
spatial dependence, we obtain $\nabla\rightarrow\nabla
+e\vec{\bm{E}}\partial_\varepsilon$ \cite{Mishchenko}, which gives
the general recipe of a quasi-chemical potential to translate
spatial inhomogeneities to internal field fluctuations and vice
versa. It is this connection and not the conservation of spins
that establishes the close relationship between the field-induced
spin accumulation and a field-independent spin current, which
contrary to Eq.~(\ref{zeroH}) vanishes in the stationary regime.
The spin-Hall current in Eq.~(\ref{zerowir}) is interpreted as the
response to the initial time evolution of the spin accumulation
after the electric field is switched on. When the spin
accumulation reaches its stationary value (cf. Fig.~2), the
related spin-Hall current component disappears.

Starting from the physically motivated definition of the spin
current in Eq.~(\ref{spinc}) and using the analytic solution in
Eq.~(\ref{mvfkE}) derived in the Appendix, we obtain the following
general result for the Laplace transformed spin-Hall current
\begin{equation}
j_y^z(s)=\frac{eE\tau}{\hbar}\left(\frac{\hbar
K}{m}\right)^2s\tau\sum\limits_{\bm{k}}n(\varepsilon_{\bm{k}})
\frac{4\omega_{\bm{k}}^4(1+2s\tau)+2\sigma^2\omega_{\bm{k}}^2(1+3s\tau)-\sigma^4s\tau}
{\left[\sigma^2 s\tau+2\omega_{\bm{k}}(2s\tau +1)\right]^2%
\left[\sigma^2s\tau+4\omega_{\bm{k}}^2(s\tau+1)\right]},\label{general1}
\end{equation}
which differs from all previous results in many respects. First,
the denominator is composed of three factors, the zeros of which
characterize spin eigenmodes. The in-plane spin precision is
characterized by poles resulting from
$1/\left[\sigma^2s\tau+4\omega_{\bm{k}}^2(s\tau+1)\right]^2$.
Resonances of this kind appear both in the spin accumulation and
in the charge current. In addition, there is the factor
$\sigma^2s\tau+4\omega_{\bm{k}}^2(s\tau+1)$ in the denominator of
Eq.~(\ref{general1}) that is associated with out-of-plane spin
eigenmodes. The most striking discrepancy results from the
$\bm{k}$ sum over the entire spin-orbit coupled Fermi sea, whereas
only contributions from the Fermi surface are needed in charge
transport problems. All states contribute to the time-dependent
spin-Hall current. This astonishing result refers both to the spin
current in [Eq.~(\ref{zerowir})] and in [Eq.~(\ref{general1})].

\begin{floatingfigure}{7.0cm}
\centerline{\includegraphics*[width=7.7cm]{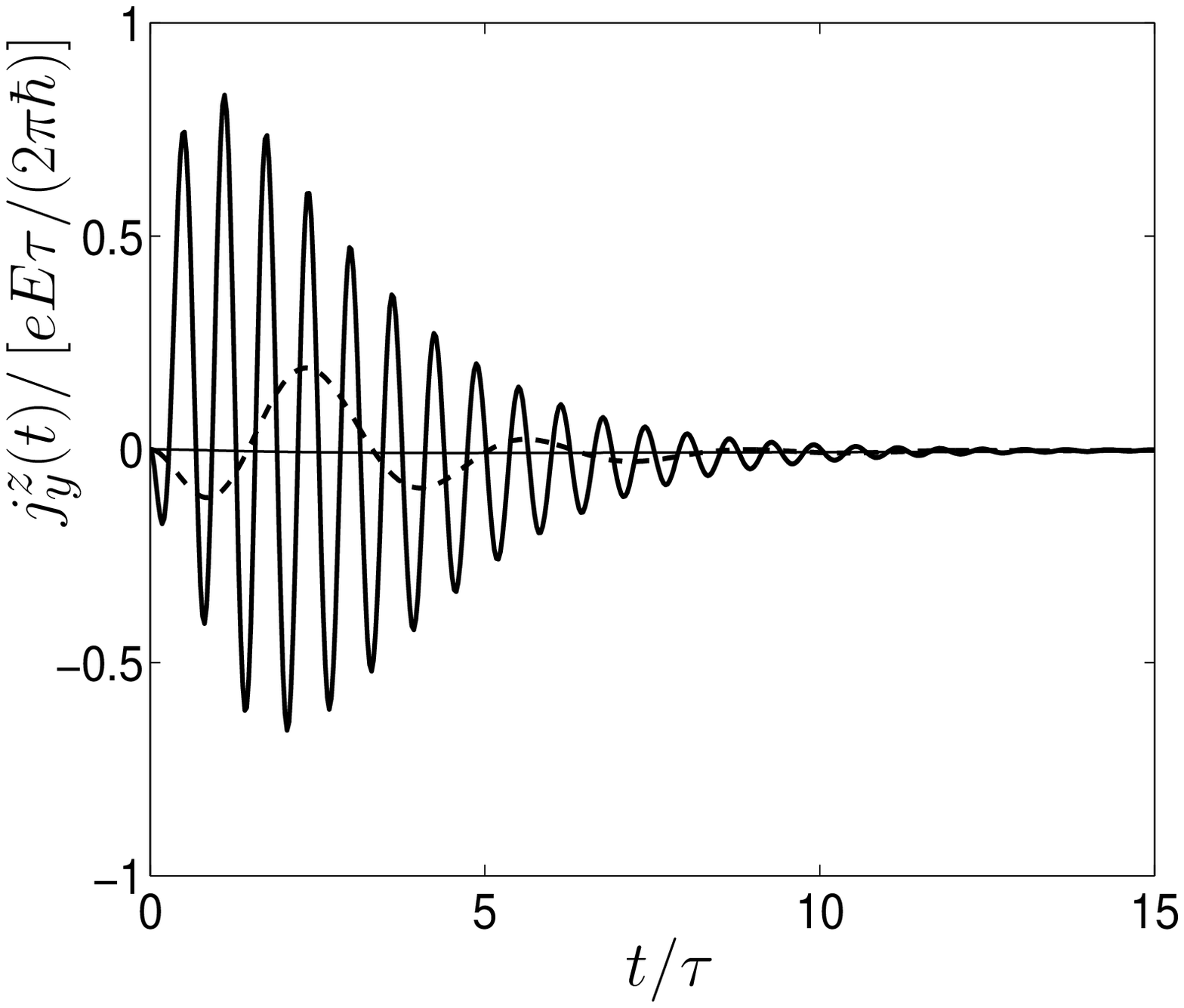}}%
\fcaption{Time dependence of the normalized spin-Hall current for
$\omega_{k_F}\tau =5$ (thick solid line), 1 (thick dashed line),
and 0.1 (thin solid line) at $T=0$.\label{abb0}}
\end{floatingfigure}
\narrowtext To clarify the origin of this peculiarity, we note
that the spin current is not due to displacements of carriers but
induced by the change of the magnetic moment. This situation is
quite similar to the diamagnetism in normal metals, which is also
determined by all states in the entire Brillouin zone
\cite{Abrikosov}. Finally, we point out that the $\bm{k}$ integral
in Eq.~(\ref{general1}) leads to a logarithmic singularity, the
due treatment of which requires a careful consideration of the
kinetic equations in the limit $\omega_{\bm{k}}\rightarrow 0$.

Applying an integration by parts, we obtain an equivalent
expression for the spin-Hall current, in \widetext which the
logarithmic contributions are singled out and in which at zero
temperature the $\bm{k}$ integral is artificially fixed at the
Fermi surface:
\begin{eqnarray}
j_y^z(s)&=&-\frac{eE\tau}{2\hbar}\frac{\hbar^2}{m}
\sum\limits_{\bm{k}}n^{\prime}(\varepsilon_{\bm{k}})\biggl\{%
\frac{2\omega_{\bm{k}}^2(2s\tau+1)}
{\left[\sigma^2s\tau+2\omega_{\bm{k}}^2(2s\tau+1)\right]}\label{gen22}\\
&+&\frac{(1+s\tau)(8s\tau+3)}{2(2s\tau+1)}\ln%
\left[1+\frac{2\omega_{\bm{k}}^2(2s\tau+1)}{\sigma^2s\tau}\right]%
-\frac{8(s\tau)^2+15s\tau+6}{4(s\tau+1)}\ln%
\left[1+\frac{4\omega_{\bm{k}}^2(s\tau+1)}{\sigma^2s\tau}\right]%
\biggl\}.\nonumber%
\end{eqnarray}
From this equation, it is concluded that in the steady state
($s\rightarrow 0$) the spin-Hall current vanishes for the impure
Rashba model of two-dimensional electrons. This finding agrees
completely with former conclusions derived from the "conventional"
definition of the spin-Hall current \cite{Mishchenko}.

Applying an inverse Laplace transformation, the time evolution of
$j_y^z$ can be studied. Fig.~3 shows an example for the
time-dependence of the spin-Hall current, which is induced by
switching on the electric field at $t=0$. Depending on the
coupling parameter $\omega_{k_F}\tau$, strong oscillations of the
spin-Hall current initially develop, which are completely damped
out after a couple of scattering times.

If time-dependent electric fields are applied, the spin-Hall
conductivity $\sigma_{sH}$ becomes nonzero. The
frequency-dependent spin-Hall conductivity is obtained from
Eq.~(\ref{gen22}) by an analytic continuation ($s\rightarrow
-i\omega$). Numerical results for the real and imaginary part of
$\sigma_{sH}$ are shown in Fig.~4 and 5 by thick solid lines. We
focus on the zero-temperature case and compare with previous
results obtained from the "conventional" definition of the
spin-Hall current \cite{Mishchenko} [dashed lines as calculated
from Eq.~(\ref{Halperin})].
\begin{figure}[!b]
\begin{minipage}[t]{7.5cm}
\centerline{\includegraphics*[width=7.5cm]{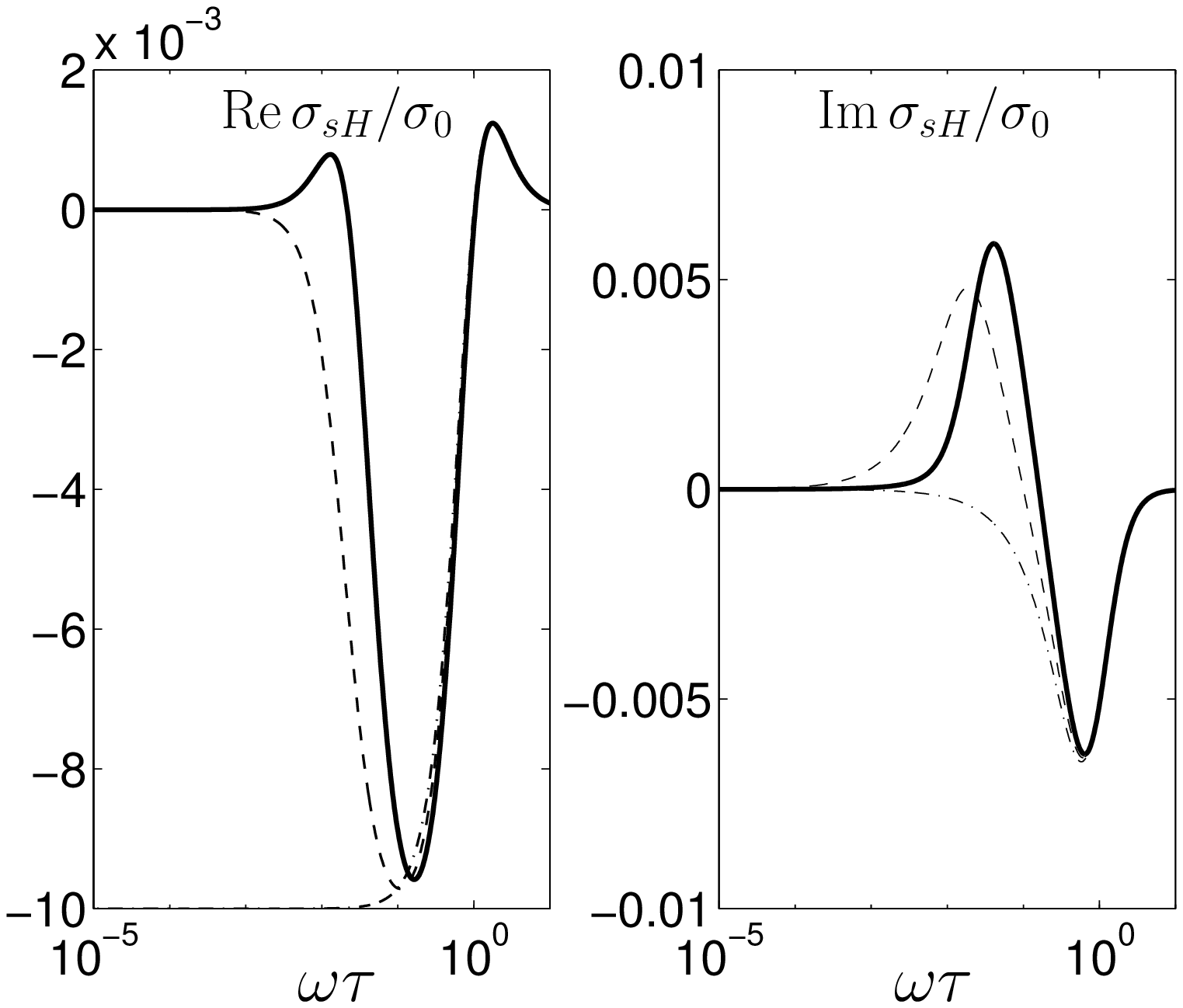}}
\fcaption{Semilogarithmic plot of the real and imaginary parts of
the normalized spin-Hall conductivity as a function of
$\omega\tau$ for $\omega_{k_F}\tau=0.1$ (thick solid line). The
dotted (dash-dotted) line is calculated from Eq.~(\ref{Halperin})
[Eq.~(\ref{so1})]. $\sigma_0$ is given by $e/(2\pi\hbar)$.
\label{abb3}}
\end{minipage}
\hfill
\begin{minipage}[t]{7.5cm}
\centerline{\includegraphics*[width=7.5cm]{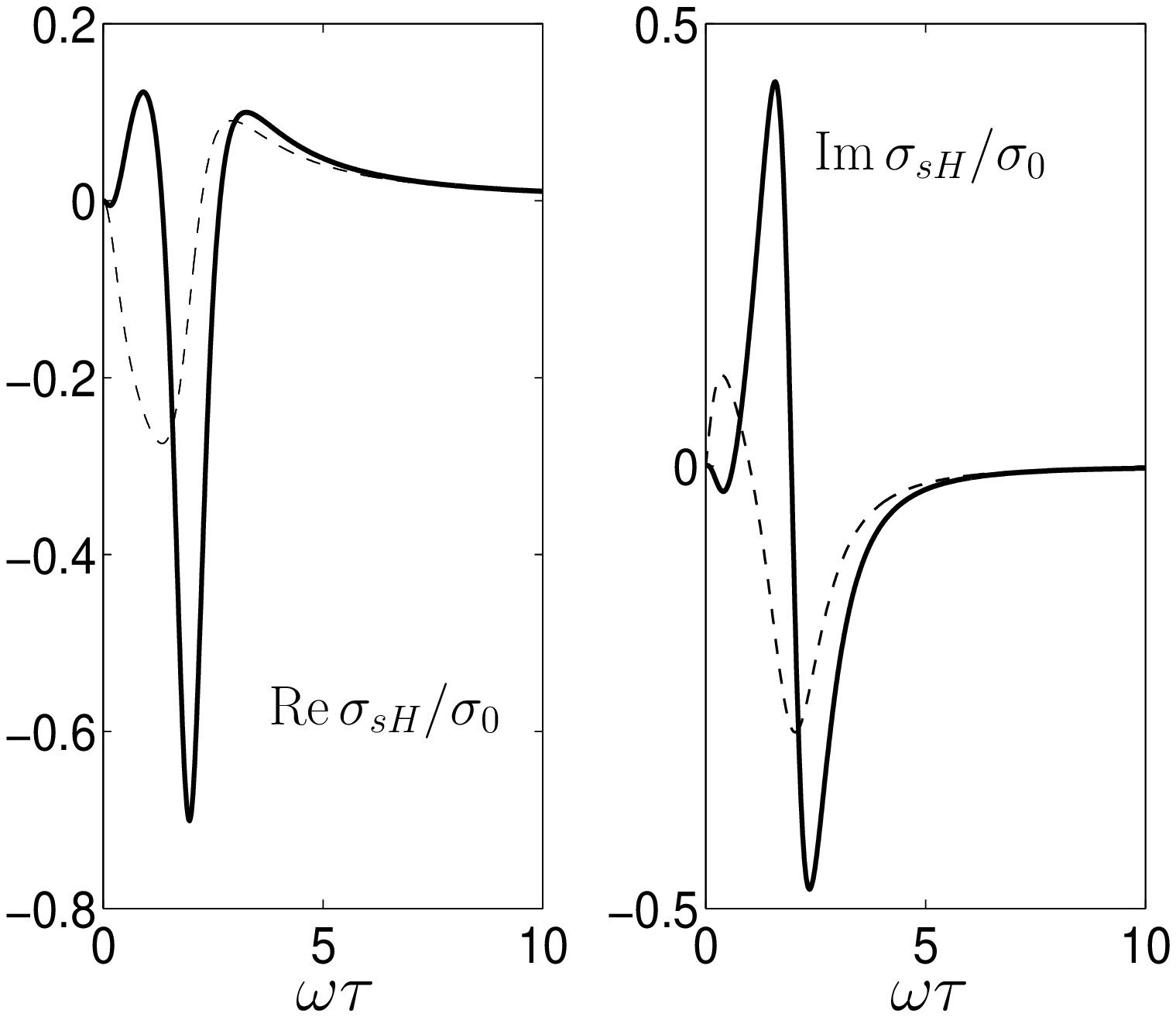}} \fcaption{Real
and imaginary parts of the normalized spin-Hall conductivity as a
function of $\omega\tau$ for $\omega_{k_F}\tau=1$ (thick solid
line). The dashed line is calculated from Eq.~(\ref{Halperin}).
\label{abb4}}
\end{minipage}
\end{figure}
Both approaches predict a sharp resonance in the ac spin-Hall
conductivity, when the condition $\omega_{k_F}\tau \gg1$ is
satisfied. The enhancement of the spin-Hall current appears at
$\omega=2\omega_{k_F}$. As seen from Fig.~4, remnants of this
feature survive even in the case $\omega_{k_F}\tau\lesssim 1$. In
the limit $\omega_{k_F}\tau\ll 1$, both approaches agree and
result in
\begin{equation}
j_y^z(\omega)=-\frac{eE}{2\pi\hbar}\left(\frac{\omega_{k_F}\tau}{1-i\omega\tau}\right)^2,
\label{so1}
\end{equation}
which is plotted by the dash-dotted line in Fig.~4. From
Eq.~(\ref{so1}), the non-analytic behavior of the spin-Hall
current becomes obvious because the approximation fails in
predicting zero spin-Hall current in the limit ($\omega
\rightarrow0$), which requires the general result in
Eq.~(\ref{general1}).

Comparing the solid and dashed lines in Figs.~4 and 5, it is
tempting to conclude that there is no qualitative difference
between the results of both approaches. That this conclusion is
only partly true shows a consideration of clean samples
($\tau\rightarrow \infty$). In this case, our approach becomes
completely exact and we obtain
\begin{equation}
j_y^z(\omega)=\frac{eE}{8\pi\hbar}\biggl\{\frac{1}{2}
\ln\left(1-\frac{4\omega_{k_F}^2}{\omega^2}\right)+
\frac{\omega^4}{(4\omega_{k_F}^2-\omega^2)^2}+
\frac{\omega^2}{2(4\omega_{k_F}^2-\omega^2)}-\frac{1}{2} \biggl\},%
\label{free1}
\end{equation}
which gives a logarithmic divergency at vanishing frequency
\begin{equation}
j_y^z(\omega\rightarrow 0)=-\frac{eE}{8\pi\hbar}\ln
\frac{\omega}{2\omega_{k_F}}.\label{log}
\end{equation}
Most previous approaches predict in this case a finite universal
spin-Hall conductivity $\sigma_{sH}=-e/(8\pi\hbar)$
\cite{Murakami,Sinova1}. In contrast, based on a physically
motivated definition of the spin-Hall current and on an exact
procedure, we obtain neither zero nor an universal value but a
logarithmic divergency for the spin-Hall conductivity at zero
frequency.

Unfortunately, recent calculations \cite{PZhang} of the physical
spin-Hall conductivity in the clean limit of the non-interacting
Rashba model do not agree with our exact result in
Eq.~(\ref{log}). The authors started from the same physically
motivated definition of the spin-Hall current (Eq.~(5) in Ref.
\cite{PZhang}) and calculated $j_y^z$ via the sourceless
continuity equation. However, strictly speaking, the spin current,
which is directly calculated from the $\bm{\kappa}$ derivative of
the averaged distribution function
$\overline{\vec{\bm{f}}}(\bm{k},\bm{\kappa}\mid t)$, is not
related to source but to vortex fields. To illustrate the
situation, let us treat the set of kinetic equations for
$\vec{\bm{E}}=\bm{0}$ and to first-order in $\bm{\kappa}$. These
equations contain two curl contributions namely $\sim
[\bm{K}\times\bm{\kappa}]\overline{f}(\bm{k},\bm{\kappa}\mid s)$
and $\sim
[\bm{\kappa}\times\overline{\vec{\bm{f}}}(\bm{k},\bm{\kappa}\mid
s)]$, which do not enter any continuity equation. The first vector
describes the coupling between charge and spin degrees of freedom
and gives rise to the field-independent spin-Hall current. If an
electric field is switched on, the replacement
$\bm{\kappa}\rightarrow\bm{\kappa}-ie\vec{\bm{E}}\partial_{\varepsilon}$
generates a contribution $\sim
-ie[\bm{K}\times\vec{\bm{E}}]\partial_{\varepsilon}\overline{f}(\bm{k},\bm{\kappa}\mid
s)$, which is responsible for the appearance of the field-induced
spin accumulation. For the mean values
$\vec{\bm{f}}(\varepsilon\mid
s)=\overline{\vec{\bm{f}}}(\bm{k},\bm{\kappa}\mid
s)\mid_{\bm{\kappa}=\bm{0}}$ and $\widehat{j}(\varepsilon\mid
s)=is\overline{\partial_{\bm{\kappa}}\otimes\vec{\bm{f}}(\bm{k},\bm{\kappa}\mid
s)}\mid_{\bm{\kappa}=\bm{0}}/2$, which are used to express the
spin accumulation and the field-independent spin-Hall current,
respectively, we obtain a relationship of the form
\begin{equation}
\frac{s}{2}f^i(\varepsilon\mid s)n=eE_jj_j^i(\varepsilon\mid
s)n^{\prime}, \label{ende}
\end{equation}
which is confirmed by our approach [compare Eq.~(\ref{Edel}) and
Eq.~(\ref{zerowir})]. This equation is not valid for the
"conventional" field-independent spin-Hall current in
Eq.~(\ref{zeroH}). The interrelation in Eq.~(\ref{ende}) is based
on the assumption that the field effects can be accounted for by a
quasi-chemical potential in the electronic part of the density
matrix. For $\alpha=0$, it is known that this supposition leads to
the Einstein relation between the mobility and the diffusion
coefficient. Recently, this hypothesis has also been accepted for
systems with spin-orbit interaction
\cite{Mishchenko,Burkov1,Blei04}, although its application becomes
more subtle due to the spin splitting of the energy bands. An
analogous use of quasi-chemical potentials for the treatment of
field effects on the spin density requires further justification.

The appearance of the logarithmic dependence of the ac spin-Hall
conductivity in the disordered two-dimensional Rashba model remind
us very much of the logarithmic quantum corrections in the theory
of weak localization \cite{Lee}. Based on scaling arguments, it
was shown that the ac conductivity of a disordered two-dimensional
electron gas exhibits a logarithmic divergency in the
zero-frequency limit [$\sigma\sim\ln(\omega\tau)$]. Our result for
the spin-Hall conductivity in Eq.~(\ref{log}) is comparable to
this dependence observed in the completely other field of weak
localization.

\section{Summary}
Based on the kinetic equations for the spin-density matrix of the
two-dimensional Rashba model, we treated the
electric-field-induced spin accumulation and spin transport in the
linear response regime. At zero temperature, the frequency
dependence of both the spin accumulation and the longitudinal
charge-carrier transport exhibit a sharp resonance at
$\omega=2\omega_{k_F}$ ($\omega_{k_F}=\alpha k_F/\hbar$, with
$\alpha$ being the spin-orbit coupling constant), which is due to
eigenmodes of spin excitations. The measurement of this resonance
should be possible. It allows the determination of the Rashba
coupling constant $\alpha$. Similar resonances are expected to
appear also in the $k$-cubed Rashba model for 2D holes and the
Luttinger model for 3D holes.

For the charge-carrier transport, there are two completely
equivalent procedures to calculate the conductance. The same
results are obtained by starting either from the symmetrized
product of the density and velocity operators or the time
derivative of the dipole moment. Unfortunately, this equivalence
no longer holds for the spin transport. The spin is not a
conserved quantity. Therefore, it is a serious problem to chose a
proper definition for the spin current that does not lose its
physical foundation. Most researchers preferred a definition of
the spin-Hall conductivity, which revealed rather unconventional
properties so that serious doubts arose on its experimental
relevance. Recently, a physically motivated definition of the spin
current has been suggested \cite{PZhang} that resolved a number of
difficulties of former approaches. It has been argued that the
proper effective spin current is inevitably defined as the time
derivative of the spin displacement. Applying this definition,
results for the spin-Hall conductivity are obtained that are
drastically different from previous findings. First, the approach
predicts a field-independent spin-Hall current that reflects the
initial variation of the spin accumulation after the electric
field is switched on. Contrary to previous results, this specific
spin current contribution disappears in the steady state. Its
physical origin is due to the initial time evolution of the spin
polarization. Furthermore, for a clean two-dimensional Rashba
model, we obtain a spin-Hall conductivity that exhibits a
logarithmic dependence at low-frequencies and not an universal
constant value. This observation, which is an exact result,
reminds us on the well-known frequency-dependent conductivity of a
disordered two-dimensional system in the theory of weak
localization.

\begin{acknowledgments}
Partial financial support by the Deutsche Forschungsgemeinschaft
and the Russian Foundation of Basic Research under the grant
number 05-02-04004 is gratefully acknowledged.
\end{acknowledgments}
\appendix
\section{Solution of the kinetic equations}
The kinetic Eqs.~(\ref{kin1}) and (\ref{kin2}) are solved by an
perturbational approach with respect to $\vec{\bm{E}}$ and
$\vec{\bm{\kappa}}$. This calculation exploits the formal exact
solutions of these equations given by
\begin{equation}
\vec{\bm{f}}=\frac{\sigma\bm{r}-2\,
\vec{\bm{\omega}}_{\bm{k}}\times\bm{r} +4\,
\vec{\bm{\omega}}_{\bm{k}}(\vec{\bm{\omega}}_{\bm{k}}\cdot
\bm{r})/\sigma}{\sigma^2+4\omega_{\bm{k}}^2}, \label{los1}
\end{equation}
with $\sigma=s+1/\tau$,
$\bm{r}=\bm{R}+\overline{\vec{\bm{f}}}/\tau$ and
\begin{eqnarray}
&&\bm{R}=\frac{i\hbar}{m}(\bm{\kappa}\cdot\bm{k})\vec{\bm{f}}
-\frac{i\hbar}{m}(\bm{K}\times\bm{\kappa})f-\frac{e\vec{\bm{
E}}}{\hbar}\nabla_{\bm{k}}\vec{\bm{f}}\nonumber\\
&&+\frac{1}{\tau} \frac{\partial}{\partial\varepsilon_{\bm{k}}}
\overline{f\hbar\vec{\bm{\omega}}_{\bm{k}}}-\frac{\hbar\vec{\bm{\omega}}_{\bm{k}}}{\tau}
\frac{\partial}{\partial\varepsilon_{\bm{k}}} \overline{f}.
\label{eqR}
\end{eqnarray}
For the angle-averaged spin-density matrix, we obtain
\begin{equation}
\overline{\vec{\bm{f}}}=\sigma\tau \frac{\sigma
\overline{\bm{R}}-2\,\overline{\vec{\bm{\omega}}_{\bm{k}}\times\bm{R}}
+4\,\overline{\vec{\bm{\omega}}_{\bm{k}}(\vec{\bm{\omega}}_{\bm{k}}\cdot
\bm{R}})/\sigma}{\sigma^2s\tau+2\omega_{\bm{k}}^2(2s\tau+1)},
\label{los2}
\end{equation}
for their $x,\, y$ components, while for the $z$ component it
follows
\begin{equation}
\overline{\vec{\bm{f}}^z}=\sigma\tau \frac{\sigma
\overline{\bm{R}}-2\,\overline{\vec{\bm{\omega}}_{\bm{k}}\times\bm{R}}}%
{\sigma^2s\tau+4\omega_{\bm{k}}^2}.
\label{los3}
\end{equation}
The lowest-order solutions in $\bm{E}=\bm{0}$ and
$\bm{\kappa}=\bm{0}$ (the corresponding elements of the density
matrix are denoted by $f_{00}$ and $\vec{\bm{f}}_{00}$) are
easiliy obtained
\begin{equation}
f_{00}=\frac{n(\varepsilon_{\bm{k}})}{s}=\overline{f_{00}},\quad
\vec{\bm{f}}_{00}=-\hbar\vec{\bm{\omega}}_{\bm{k}}\frac{n^{\prime}}{s},\quad
\overline{\vec{\bm{f}}_{00}}=\bm{0}, \label{f00}
\end{equation}
where $n^{\prime}$ denotes the derivative with respect to
$\varepsilon_{\bm{k}}$. Both quantities $f_{00}$ and
$\vec{\bm{f}}_{00}$ do not depend on time and are therefore
conserved under the condition of thermodynamic equilibrium. For
the derivation of this result it was necessary to consider the
spin-orbit coupling in the collision integral [the second and
third term on the right-hand side of Eq.~(\ref{kin2})]. Next, let
us calculate the lowest-order correction due to the electric field
($\bm{E}\ne \bm{0}$, $\bm{\kappa}=\bm{0}$). Taking into account
Eqs.~(\ref{kin1}) and (\ref{kin2}) together with Eq.~(\ref{los2}),
we obtain
\begin{equation}
f_{0\bm{E}}=-\frac{eE}{\sigma s}\frac{\hbar k_x}{m}n^{\prime},
\quad \overline{f_{0\bm{E}}}=0 ,\label{f0E}
\end{equation}
\begin{equation}
\vec{\bm{f}}_{0\bm{E}}=%
\vec{\bm{\omega}}_{\bm{k}}(eEk_x)\frac{\hbar^2n^{\prime\prime}}{m\sigma
s}+n^{\prime}\tau
\frac{\sigma\vec{\bm{\omega}}_{\bm{E}}-%
2\vec{\bm{\omega}}_{\bm{k}}\times\vec{\bm{\omega}}_{\bm{E}}%
+4(\vec{\bm{\omega}}_{\bm{k}}\cdot\vec{\bm{\omega}}_{\bm{E}})\vec{\bm{\omega}}_{\bm{k}}/\sigma}%
{\left[\sigma^2s\tau+2\omega_{\bm{k}}(2s\tau+1)\right]}
,\label{vf0E}
\end{equation}
\begin{equation}
\overline{\vec{\bm{f}}}_{0\bm{E}}=\frac{\vec{\bm{\omega}}_{\bm{E}}}{\sigma
s}\biggl\{(\varepsilon_{\bm{k}}n^{\prime})^{\prime}
-\frac{2\sigma\tau\omega_{\bm{k}}^2n^{\prime}}{\sigma^2s\tau+2\omega_{\bm{k}}^2(2s\tau+1)}\biggl\}
, \label{mvf0E}
\end{equation}
where we used the abbreviation $\vec{\bm{\omega}}_{\bm{E}}=e\hbar
(\bm{K}\times\vec{\bm{E}})/m$. Next, we calculate
$f_{\bm{\kappa}0}$ and $\vec{\bm{f}}_{\bm{\kappa}0}$ by collecting
the first-order contributions in $\bm{\kappa}$ and by setting
$\vec{\bm{E}}=\bm{0}$. Neglecting corrections, which are of the
order $\hbar^2K/m\varepsilon_{\bm{k}}$, we obtain
\begin{equation}
f_{\bm{\kappa}0}=\frac{i\hbar}{m\sigma
s}(\bm{\kappa}\cdot\bm{k})n,%
\label{fk0}
\end{equation}
\begin{equation}
\vec{\bm{f}}_{\bm{\kappa}0}=
-i\vec{\bm{\omega}}_{\bm{k}}(\bm{k}\cdot\bm{\kappa})\frac{\hbar^2n^{\prime}}{m\sigma
s}-in\tau
\frac{\sigma\vec{\bm{\omega}}_{\bm{\kappa}}-%
2\vec{\bm{\omega}}_{\bm{k}}\times\vec{\bm{\omega}}_{\bm{\kappa}}%
+4(\vec{\bm{\omega}}_{\bm{k}}\cdot\vec{\bm{\omega}}_{\bm{\kappa}})\vec{\bm{\omega}}_{\bm{k}}/\sigma}%
{\left[\sigma^2s\tau+2\omega_{\bm{k}}(2s\tau+1)\right]}
\label{vfk0}
\end{equation}
\begin{equation}
\overline{\vec{\bm{f}}}_{\bm{\kappa}0}=%
-\frac{i\vec{\bm{\omega}}_{\bm{\kappa}}}{\sigma
s}\biggl\{(\varepsilon_{\bm{k}}n)^{\prime}
-\frac{2\sigma\tau\omega_{\bm{k}}^2n}{\sigma^2s\tau+2\omega_{\bm{k}}^2(2s\tau+1)}\biggl\}%
. \label{mvfk0}
\end{equation}
There is an interesting symmetry between the vectors
$\vec{\bm{f}}_{\bm{\kappa}0}$ and $\vec{\bm{f}}_{0\bm{E}}$. From
Eq.~(\ref{vfk0}), the electric-field-induced contribution in
Eq.~(\ref{vf0E}) is obtained by the replacement
$\bm{\kappa}\rightarrow -ieE\vec{\bm{e}}_x\partial_{\varepsilon}$,
where the derivative refers specifically to the charge density
$n$. Finally, we need the first-order corrections in $\bm{\kappa}$
and $\bm{E}$ of the angle-averaged component of the density
matrix, which are expressed by
\begin{equation}
\overline{f}_{\bm{\kappa}\bm{E}}=
-\frac{i\kappa_x}{s}\biggl\{\frac{\hbar
K}{m}\overline{\vec{\bm{f}}^y}_{\bm{0}\bm{E}}+\frac{eE}{m\sigma
s}\left[ 2(\varepsilon n)^{\prime}-n\right]\biggl\}, \label{mfkE}
\end{equation}
\begin{equation}
{\overline{\vec{\bm{f}}^z}}_{\bm{\kappa}\bm{E}}=%
2i\kappa_y\frac{eE\tau}{\hbar}\left(\frac{\hbar K}{m}\right)^2\tau
n(\varepsilon_{\bm{k}})
\frac{4\omega_{\bm{k}}^4(1+2s\tau)+2\sigma^2\omega_{\bm{k}}^2(1+3s\tau)-\sigma^4s\tau}
{\left[\sigma^2 s\tau+2\omega_{\bm{k}}(2s\tau +1)\right]^2%
\left[\sigma^2s\tau+4\omega_{\bm{k}}^2(s\tau+1)\right]}.
\label{mvfkE}
\end{equation}

\end{document}